\documentclass[runningheads]{llncs}
\usepackage{graphicx}
\usepackage{esvect}
\usepackage{algorithm}
\usepackage{algorithmicx}
\usepackage{algpseudocode}

\usepackage{subfig}
\usepackage{amsmath}
\usepackage{hhline}
\usepackage{xkeyval,xcolor}
\makeatletter
\newlength{\sfp@hseplen}\newlength{\sfp@vseplen}
\define@cmdkey{subfigpos}[sfp@]{pos}[ul]{}
\define@cmdkey{subfigpos}[sfp@]{font}[\small]{}
\define@cmdkey{subfigpos}[sfp@]{vsep}[5pt]{\setlength{\sfp@vseplen}{\sfp@vsep}}
\define@cmdkey{subfigpos}[sfp@]{hsep}[5pt]{\setlength{\sfp@hseplen}{\sfp@hsep}}
\newcommand{\subfigimg}[3][,]{%
  \setkeys{Gin,subfigpos}{pos,font,vsep,hsep,#1}
  \setbox1=\hbox{\includegraphics{#3}}
  \ifnum\pdfstrcmp{\sfp@pos}{ul}=0
    \leavevmode\rlap{\usebox1}
    \rlap{\hspace*{\sfp@hsep}\raisebox{\dimexpr\ht1-\sfp@vsep}{\sfp@font{#2}}}
    \phantom{\usebox1}
  \else\ifnum\pdfstrcmp{\sfp@pos}{ur}=0
    \leavevmode\usebox1
    \llap{\raisebox{\dimexpr\ht1-\sfp@vsep}{\sfp@font{#2}}\hspace*{\sfp@hsep}}
  \else\ifnum\pdfstrcmp{\sfp@pos}{lr}=0
    \leavevmode\usebox1
    \llap{\raisebox{\sfp@vsep}{\sfp@font{#2}}\hspace*{\sfp@hsep}}
  \else
    \leavevmode\rlap{\usebox1}
    \rlap{\hspace*{\sfp@hseplen}\raisebox{\sfp@vsep}{\sfp@font{#2}}}
    \phantom{\usebox1}
  \fi\fi\fi
}
\makeatother

\begin{document}
\title{Deep Reinforcement Learning for\\Small Bowel Path Tracking using\\Different Types of Annotations}
\titlerunning{Deep Reinforcement Learning for Small Bowel Path Tracking}
%
\author{Seung Yeon Shin \and
Ronald M. Summers}
%
\authorrunning{S.Y. Shin et al.}
%
\institute{Imaging Biomarkers and Computer-Aided Diagnosis Laboratory, Radiology and Imaging Sciences, Clinical Center, National Institutes of Health, Bethesda, MD, USA\\
\email{\{seungyeon.shin,rms\}@nih.gov}}
\maketitle              
\begin{abstract}
Small bowel path tracking is a challenging problem considering its many folds and contact along its course. For the same reason, it is very costly to achieve the ground-truth (GT) path of the small bowel in 3D. In this work, we propose to train a deep reinforcement learning tracker using datasets with different types of annotations. Specifically, we utilize CT scans that have only GT small bowel segmentation as well as ones with the GT path. It is enabled by designing a unique environment that is compatible for both, including a reward definable even without the GT path. The performed experiments proved the validity of the proposed method. The proposed method holds a high degree of usability in this problem by being able to utilize the scans with weak annotations, and thus by possibly reducing the required annotation cost.

\end{abstract}
\section{Introduction}\label{sec:intro}
The small bowel is the longest section of the digestive tract (about 6$m$). Despite its length, it is pliable and has many folds so that it can fit into the abdominal cavity~\cite{cc19}. As a consequence, it has abundant contact between parts of itself as well as with adjacent organs such as the large bowel. Its appearance also varies dynamically along its course since bowel contents are locally highly variable.

Identifying the small bowel from a computed tomography (CT) scan is a prerequisite for inspection of its normalcy, including detection of diseases that can exist in the small bowel. To replace the manual interpretation, which is laborious and time-consuming, there have been researches on automatic small bowel segmentation~\cite{zhang13,shin20,shin21}. Considering the difficulty of labeling the small bowel, these works have focused on data-efficient methods by incorporating a shape prior~\cite{shin20} or by developing an unsupervised domain adaptation method~\cite{shin21}.

Although the segmentation could provide fundamental information for subsequent clinical tasks by separating the small bowel from the other tissues and organs, it is insufficient to identify its course precisely due to the aforementioned contact issue. To achieve this complementary information, there have been attempts to develop path tracking methods for the small bowel in recent years~\cite{oda21,harten21,shin22_spie,shin22_isbi}. While a deep orientation classifier is trained to predict the direction of the small bowel in cine-MR scans~\cite{harten21}, the 3D U-Net~\cite{cicek16} is used to estimate the distance from the small bowel centerlines for each voxel~\cite{oda21}. CT scans that contain air-inflated bowels, where the walls are more distinguishable from the lumen than with other internal material, were used in \cite{oda21}. On account of the labeling difficulty again, their networks were trained and evaluated using sparse annotation (several axial slices)~\cite{oda21}, or with a set of bowel segments (88$mm$ long on average)~\cite{harten21}. In \cite{shin22_spie}, a graph-based method that does not entail training with GT path is presented. Although its predicted path showed a high coverage on the entire course of the small bowel in routine CT scans, it is affected by the quality of the accompanying segmentation prediction.

Over the past several years, deep reinforcement learning (DRL) based approaches have been proposed to trace elongated structures in biomedical images, including axons~\cite{dai19}, the aorta~\cite{zhang18}, and coronary arteries~\cite{li21}. The RL formulation for each application was validated by comparing with a supervised learning counterpart, which predicts moving actions as output labels from independent inputs~\cite{zhang18,li21}. Despite the benefit, their reward is based on a distance to ground-truth (GT) path, thus precluding the training when the GT path is unavailable. 

\begin{figure}[t]
	\centering
	\includegraphics[width = 1\textwidth]{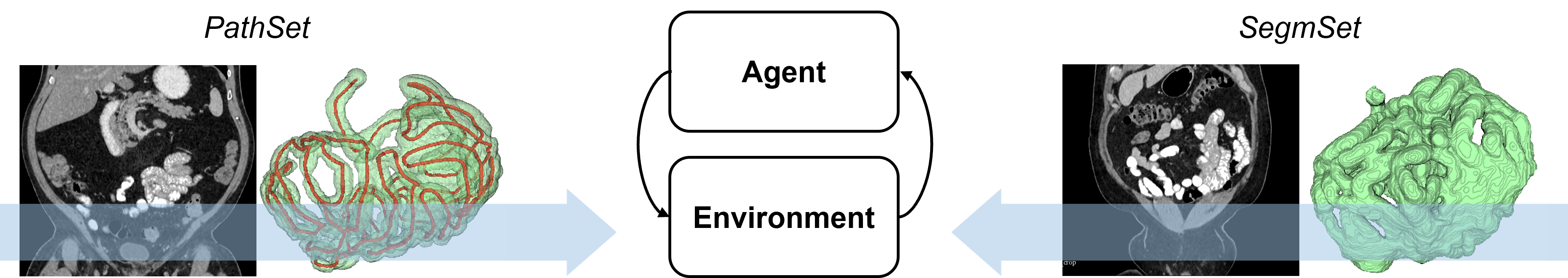}
	\caption{Main idea of this paper. We utilize scans that have only ground-truth (GT) segmentation, \emph{SegmSet}, as well as ones with GT path, \emph{PathSet}, to train a deep reinforcement learning (DRL) tracker for the small bowel. The GT path and segmentation are shown in red and green, respectively.}
	\label{fig:idea}
\end{figure}

In this paper, we propose to train a DRL tracker for the small bowel using different types of annotations. Specifically, we make use of scans that have only GT segmentation as well as ones with GT path (Figure~\ref{fig:idea}). To facilitate this, the environment our agent interacts with and the accompanying reward should not necessarily require the GT path for their definition, and should be compatible regardless of whether the GT path is available or not. We note that the annotation cost of the small bowel path is much higher than that of segmentation. Our experimental dataset will be introduced first to explain the necessity of using different types of annotations in the following section.

\section{Method}\label{sec:method}

\subsection{Dataset}
Our dataset consists of 30 intravenous and oral contrast-enhanced abdominal CT scans. Before the scan, a positive oral contrast medium, Gastrografin, was taken. All scans were done during the portal venous phase. They were cropped manually along the z-axis to include from the diaphragm through the pelvis. Then, they were resampled to have isotropic voxels of $1.5^3mm^3$.

Two types of GT labels, which are path and segmentation of the small bowel, are included in our dataset. It is separated into two subsets depending on whether or not the GT path is included. The first subset includes 10 scans that have both labels. The remaining 20 scans have only the GT segmentation. We call the subsets \emph{PathSet} and \emph{SegmSet}, respectively. An experienced radiologist used ``Segment Editor" module in 3DSlicer~\cite{fedorov12} to acquire the GTs. Our GTs include the entire small bowel from the pylorus to the ileocecal junction.

For \emph{SegmSet}, the small bowel was first drawn roughly using a large brush, and thresholded and manually fixed. This segmentation annotation took a couple of hours for each scan. The start and end coordinates of the small bowel were also recorded. For \emph{PathSet}, the GT path is drawn as an interpolated curve connecting a series of manually placed points inside the bowel. This is exceptionally time-consuming and took one full day for each scan. For these scans that have the GT path, segmentation can be acquired more easily by growing the path. A thresholding and manual correction were done too. Our work is originated from the difficulty of achieving the GT path in a larger scale. It would be nice if we can utilize the scans that are without the GT path for training a better tracker. \emph{PathSet} is used for cross validation while \emph{SegmSet} is used for training. We note that our GT path is all connected for the entire small bowel. Compared with the sparse~\cite{oda21} and segmented~\cite{harten21} ones from the previous works, ours are more appropriate to see the tracking capability for the entire course.

\subsection{Environment}\label{subsec:env}
In RL, an agent learns a policy $\pi$ from episodes generated by interacting with an environment. We use one image per episode. An agent (tracker) is initialized at a position within the small bowel, and moves within the image until certain conditions are met. The conditions are: 1) finding one end of the small bowel, 2) leaving the image, or 3) being at the maximum time step $T$. One more termination condition of zero movement is used in test time. At every time step $t$, the agent performs an action $a_t$ (movement), which is predicted by our actor network, and receives a reward $r_t$ and a new state $s_{t+1}$ as result. An episode, which is a sequence of $(s_t, a_t, r_t)$, is generated by iterating this. Each of the state space, action space, and reward will be explained in the following sections.

\begin{figure}[t]
	\centering
	\includegraphics[width = 1\textwidth]{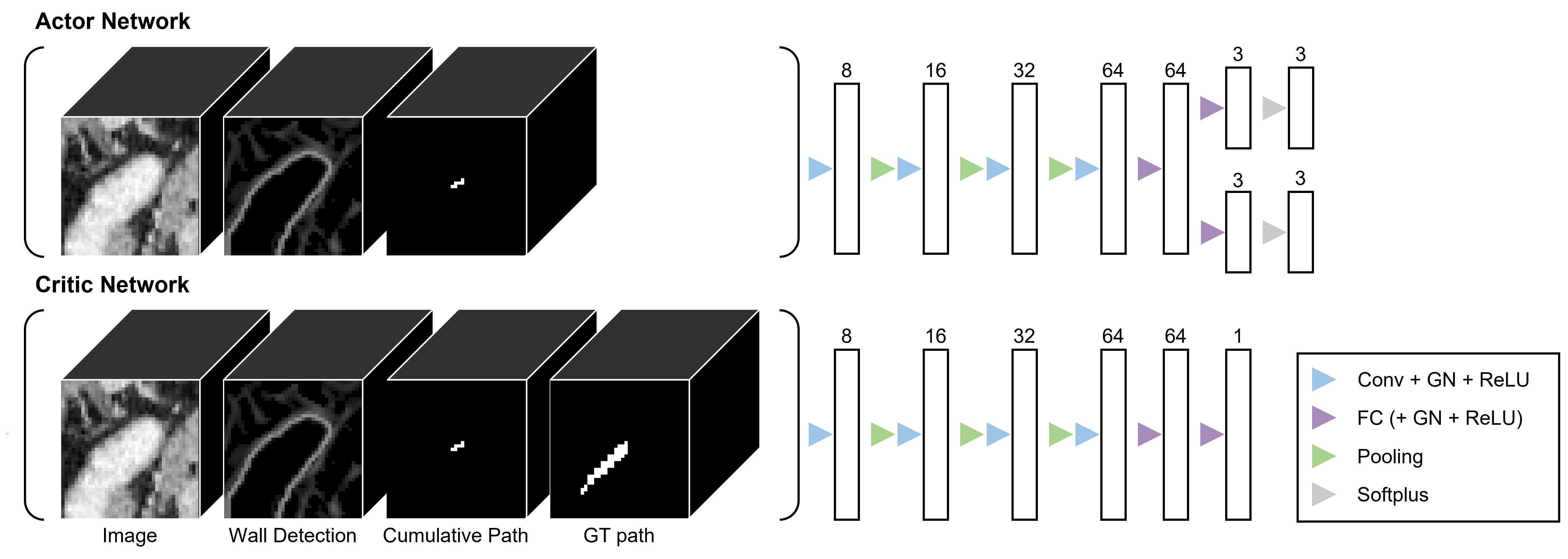}
	\caption{Architectures of our networks and their input. Boxes on the right represent feature maps. The number of channels (dimensions) is denoted on top of each feature map. GN represents the group normalization~\cite{wu18}. All convolution layers have 3x3x3 kernels. All fully connected (FC) layers are followed by a GN and a ReLU excepting the very last ones in each network.
	}
	\label{fig:network}
\end{figure}

\subsubsection{State}
At each time step $t$, local image patches, of size $60^3mm^3$, centered at the current position $p_t$ is used to represent the agent's state. They are the input to our networks. Figure~\ref{fig:network} shows example input patches. Due to the contact issue, preventing the tracker from penetrating the bowel walls is critical in this problem. To better provide this awareness to the agent, we use the wall detection as input in addition to an `ordinary' image patch. In our dataset, the lumen appears brighter than the walls due to the oral contrast. We detect the walls by finding valleys in an input volume using the Meijering filter~\cite{meijering04} as in \cite{shin22_spie}. We also provide a local patch of the agent's cumulative path in the current episode as in \cite{dai19}. It denotes the previously visited voxels around the current position, which is binary.

In this work, we use an actor-critic algorithm~\cite{konda99}. The actor decides which action to take (policy), and the critic judges how good the actor's action was (value function). They are learned simultaneously, and the critic helps reduce the variance of the policy updates. The critic is used only during training. In DRL, the policy and value function are approximated by neural networks. To this end, we use separate networks, namely, actor and critic networks. Especially, we use the asymmetric actor-critic algorithm~\cite{pinto17}. An additional information of the GT path is provided only to the critic network. It helps training the critic network and thus allows for better updates for the actor~\cite{pinto17,dai19}. All the explained input patches are concatenated and fed into the networks. For \emph{SegmSet}, the GT path is not available, and a zero tensor is used instead.

\subsubsection{Action}
The action is chosen based on the output of the actor network, which are three pairs of the parameters $(\alpha, \beta)$ of the beta distribution. The beta distribution has a finite support $[0,1]$ so that can constrain the action space more easily than using the Gaussian~\cite{chou17}. Each pair is responsible for movement along each axis. A softplus operation is applied to the output of the last fully connected layers, and then $1$ is added to ensure every $\alpha$ and $\beta$ is greater than $1$. When this condition is met, the beta distributions are unimodal. In training, a value is sampled from each beta distribution. This probabilistic sampling allows for exploration of the agent. The sampled value within $[0,1]$ is mapped to an actual displacement within $[-d_{step},d_{step}]$, and rounded off to move on the image volume. In test, the mode of the distribution, $\frac{\alpha-1}{\alpha+\beta-2}$, is deterministically used.

\subsubsection{Reward}

\begin{algorithm}[t]
	\caption{Reward for small bowel path tracking}
	\label{alg:reward}
	\begin{algorithmic}[1]
	\Require{Current agent position $p_t$, current action $a_t$, ground-truth segmentation image $I^{segm}$, wall detection image $I^{wall}$, cumulative path image $I^{cum}$, geodesic distance transform (GDT) image $I^{GDT}$, maximum GDT value achieved so far $v_{max}^{GDT}$, maximum allowed increase of GD per step $\theta$, predefined reward scale $r^{val1}$, $r^{val2}$}
    \Ensure{Reward $r_t$}
    \If{$a_t==\vv{0}$}
        \State $r_t = -r^{val1}$ \algorithmiccomment{zero movement penalty}
    \Else
	    \State $r_t = 0$
		\State Next agent position $p_{t+1} \leftarrow p_t + a_t$
		\State Set of voxels $S$ on the line between $p_t$ and $p_{t+1}$
	    \State Next GDT value $v_{t+1}^{GDT} \leftarrow I^{GDT}[p_{t+1}]$
		\If{$v_{t+1}^{GDT} > v_{max}^{GDT}$}
		    \State $\Delta v^{GDT} = v_{t+1}^{GDT} - v_{max}^{GDT}$ \;
		    \If{$\Delta v^{GDT} > \theta$}
		        \State $r_t = -r^{val2}$ \algorithmiccomment{penalty on an abrupt increase of GD}
		    \Else
		        \State $r_t = \frac{\Delta v^{GDT}}{\theta} \times r^{val2}$ \algorithmiccomment{GDT-based reward}
	        \EndIf
		    \State $v_{max}^{GDT} \leftarrow v_{t+1}^{GDT}$ \;
        \EndIf
	    \State $r_t  \mathrel{{-}{=}} (\frac{1}{|S|}{\sum_{S}{I^{wall}[s]}}) \times r^{val2}$ \algorithmiccomment{wall-based penalty}
		\If{$\sum_{S}{I^{cum}[s]} > 0$}
		    \State $r_t \mathrel{{-}{=}} r^{val1}$ \algorithmiccomment{revisiting penalty}
	    \EndIf
		\If{$I^{segm}[p_{t+1}]==False$}
		    \State $r_t = -r^{val1}$ \algorithmiccomment{out-of-segmentation penalty}
	    \EndIf
    \EndIf
    \State \Return {$r_t$}
	\end{algorithmic}
\end{algorithm}

\begin{figure}[t]
	\centering
	\begin{minipage}{0.7\textwidth}
        \subfloat[]{\includegraphics[width = 0.3\textwidth]{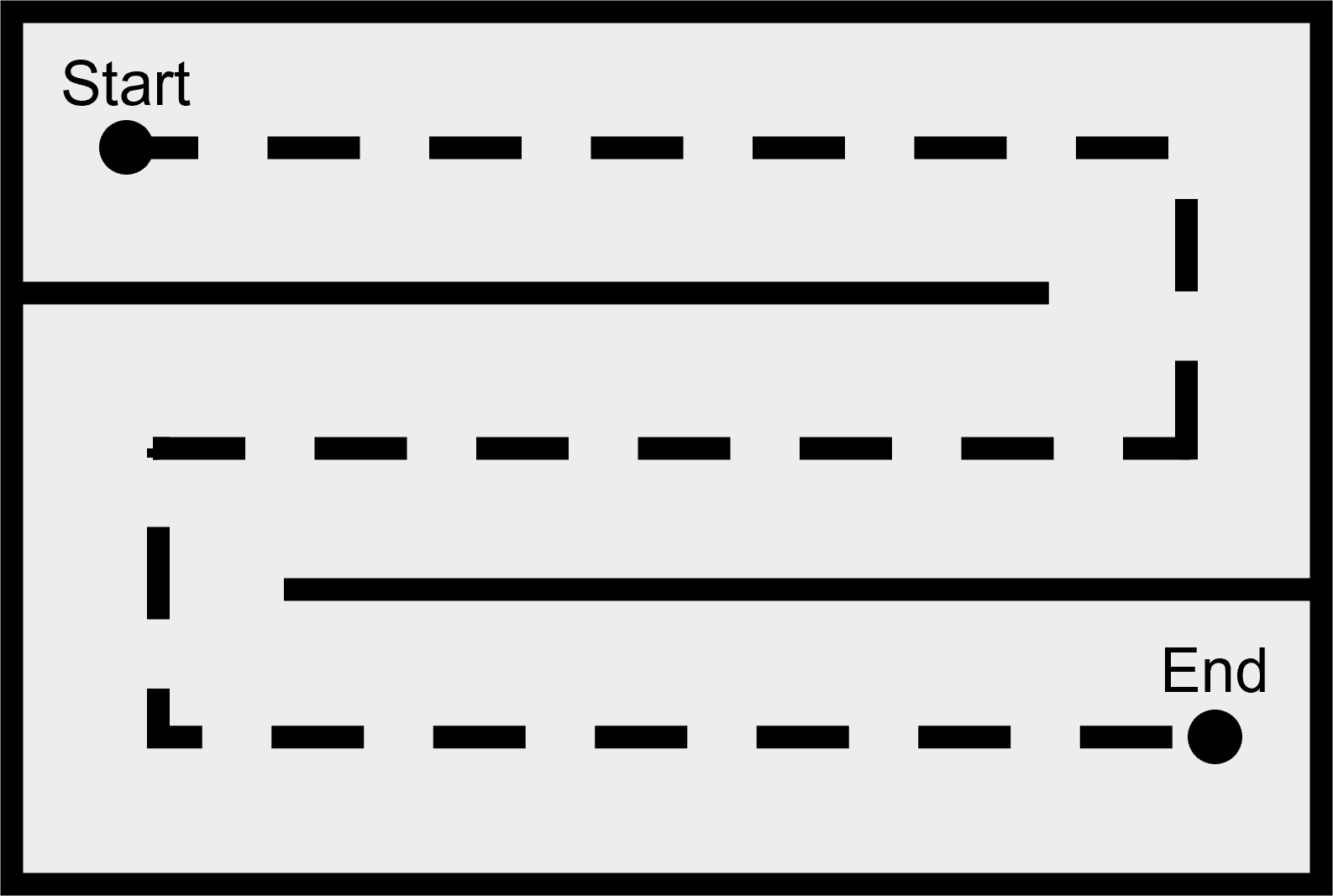}}
        \hspace{0.2cm}
	    \subfloat[]{\includegraphics[width = 0.3\textwidth]{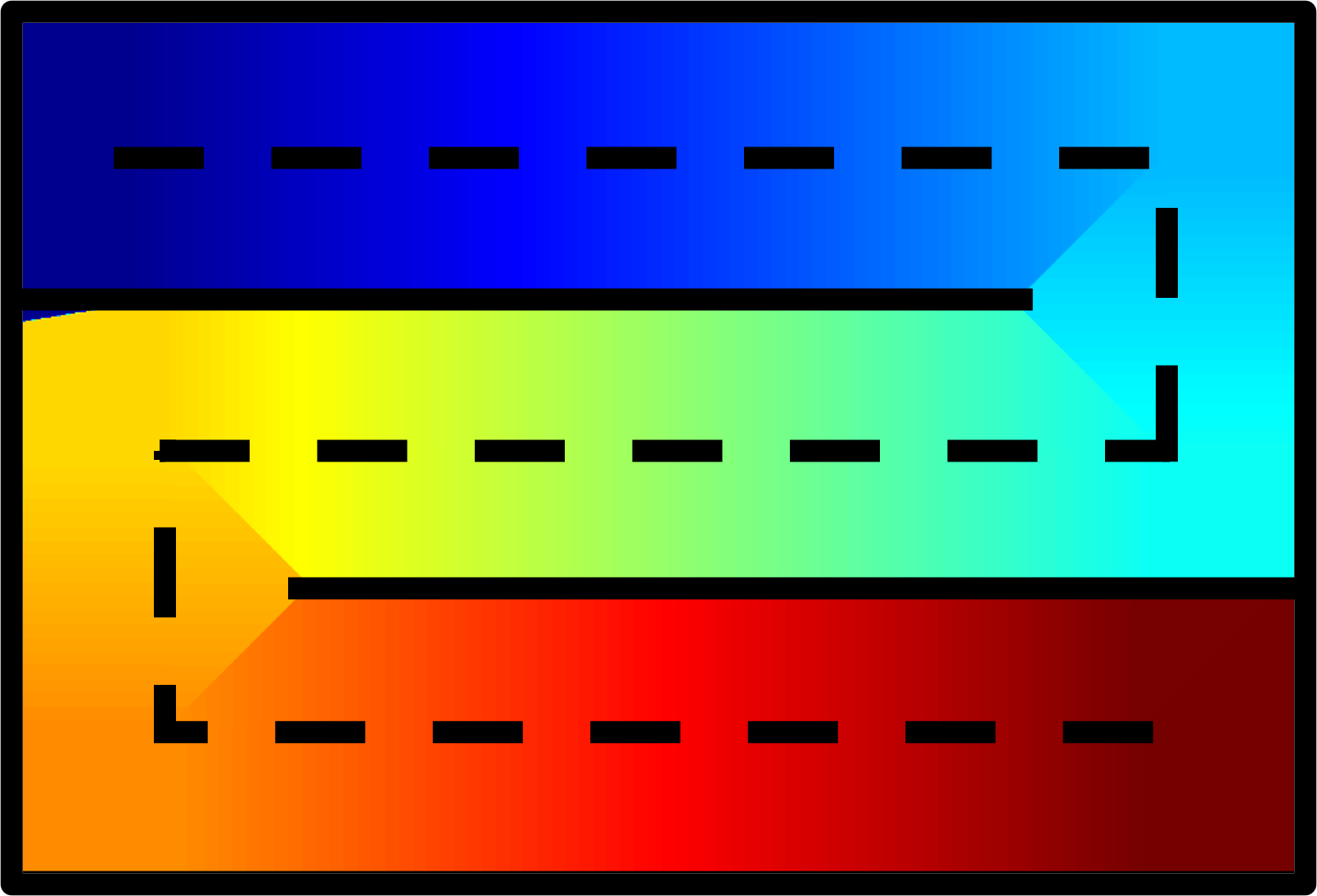}}
	    \hspace{0.2cm}
	    \subfloat[]{\includegraphics[width = 0.3\textwidth]{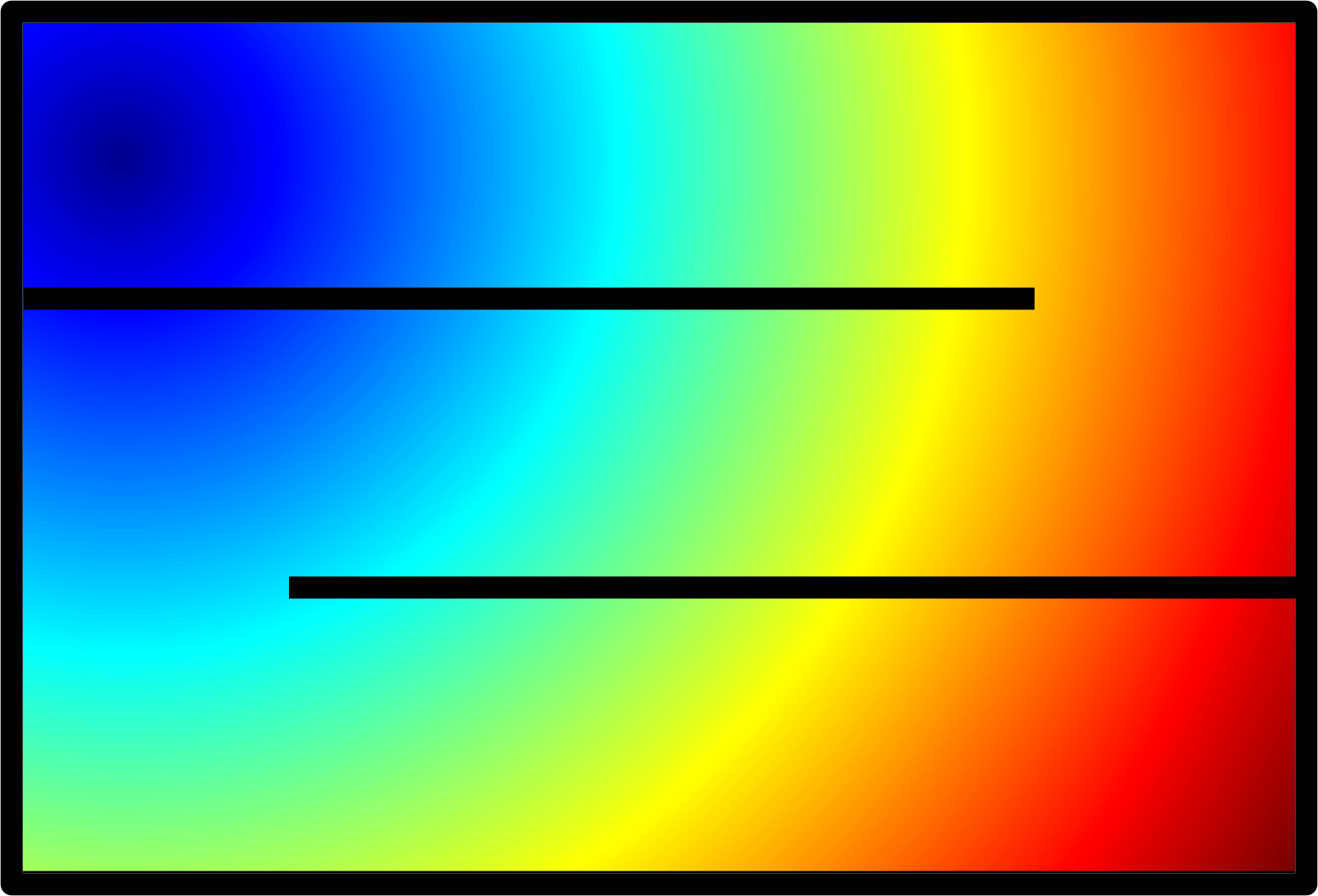}}
    \end{minipage}
	\caption{Toy examples of the geodesic distance transform (GDT), which is computed by the fast marching method~\cite{yatziv06}. (a) Toy small bowel with folds and contact. The solid and dashed lines denote the bowel wall and GT path, respectively. (b) GDT from the start when the GT path is given. The distance gradually increases from blue to red. (c) GDT when only segmentation is given.}
	\label{fig:gdt}
\end{figure}

Our reward calculation for each step is summarized in Algorithm~\ref{alg:reward}. A key point is to use the geodesic distance transform (GDT). We hypothesized the reward function based on the Euclidean distance to the closest point on the GT path, which is used in \cite{zhang18,li21}, would be inappropriate for our problem since the small bowel has local paths that are spatially close each other. It may encourage crossing the walls. The corresponding experiment will be discussed in Section~\ref{sec:results}. Figure~\ref{fig:gdt} shows toy examples of the GDT. Despite the contact issue, when the GT path is available (Figure~\ref{fig:gdt} (b)), it reflects the bowel path well. We incentivize increasing the distance from the start on the GDT. One more benefit of using the GDT is its applicability to the scans without the GT path (Figure~\ref{fig:gdt} (c)), namely \emph{SegmSet}. Being of lower quality, it still encodes the distance from the start roughly. Our objective is to utilize \emph{SegmSet} as well as \emph{PathSet} for training. To make our reward compatible for both, we do not penalize a decrease of GD. We only incentivize achieving a higher value than the maximum achieved so far. This encourages the agent to travel more to achieve a higher maximum, even though it looks to become closer to the start temporarily on the lumpy small bowel.

Another important factor is to use the bowel wall detection. We penalize passing the detected walls. Revisiting the previous path and leaving the small bowel segmentation are also penalized. For calculation of the revisiting penalty, the cumulative path is stored also as cylinders of radius $6mm$. When an episode terminates, we incentivize/penalize finally according to the coverage of the cylinderized cumulative path on the segmentation, $c$. We provide the final reward of $c \times r^{final}$ if the agent arrives at the end, otherwise $(c-1) \times r^{final}$.
In summary, our reward encourages the agent to travel to get farther from the start while not penetrating the walls, not revisiting the previous path, and not leaving the small bowel. This can be pursued even without the GT path.

\subsection{Training and Testing}
We now can make our environment using both \emph{PathSet} and \emph{SegmSet}. We run four episodes in parallel to collect training data, where two of them are from \emph{PathSet} and the other two are from \emph{SegmSet}. In training, a tracking can start from either of two ends of the small bowel, or from a middle point on the GT path with a probability of $0.3$ to diversify the episodes. Since \emph{SegmSet} has no GT path, this middle start point is chosen more carefully as a non-wall small bowel voxel. It always starts at the pylorus in test time. Each episode continues until the conditions mentioned in Section~\ref{subsec:env} are met. We used proximal policy optimization (PPO)~\cite{schulman17} to train our actor/critic networks. The detailed training algorithm can be found in supplementary material.

\subsection{Evaluation Details}
The maximum displacement along each axis per step $d_{step}$ and the maximum time step per episode $T$ were set to $10mm$ and $800$, respectively. The cell length in GDT was set to $1.5$, which is the same with the voxel size of the input volume. The maximum allowed increase of GD per step $\theta=\sqrt{3*10^2}$ was used accordingly. The reward scale $r^{val1}=4$, $r^{val2}=6$, and $r^{final}=100$ were chosen by experiments. The discount factor $\gamma=0.99$, generalized advantage estimation (GAE)~\cite{schulman16} parameter $\lambda=0.95$, clipping parameter $\epsilon=0.2$, and entropy coefficient of $0.001$ were used for PPO. A learning rate of $10^{-5}$, minibatch size of $32$, and the number of epochs $5$ were used for both networks.

We implemented our networks, including ones for the comparable methods, using PyTorch 1.8.2.
We used Adam optimizers~\cite{kingma15} for training.
We used a NVIDIA Tesla V100 32GB GPU to conduct experiments.

We provide the maximum length of the GT path that is tracked without making an error, which is used in \cite{shin22_spie}, as an evaluation metric. When comparing the predicted path with the GT path, a static distance tolerance of $10mm$ was used. We note that this metric is more important than the others used in \cite{shin22_spie}, including the precision, recall, and curve-to-curve distance, since they are computed in disregard of the order of the tracking, and the error of crossing the walls. Instead, we provide more information for the maximum length metric.

\section{Results}\label{sec:results}
\subsection{Quantitative Evaluation}

\begin{table}[t]
\centering
\caption{Quantitative comparison of different methods. Statistics on the maximum length of the GT path that is tracked without making an error are presented in $mm$. $N^{th}$ denotes the $N^{th}$ percentile. Each method is categorized into one of the supervised learning (SL), RL, and graph based methods. The first three and the next three are ones that use \emph{PathSet} or \emph{SegmSet} for training, respectively. The last two are trained using both sets. The numbers of scans used for training are shown. All ten scans in \emph{PathSet} were evaluated using a 2-fold cross validation except for `TSP~\cite{shin22_spie}' and `Ours (s) (0/20)', which are trained using only \emph{SegmSet}, thus do not need the cross validation. `0/25' denotes using additional five scans from \emph{PathSet} as \emph{SegmSet} to test the remaining five. Refer to the text for the explanation on each method.}\label{tab:quan_res}
\begin{scriptsize}
\begin{tabular}{c|c|c|c|c|c|c|c|c}
Method & CAT & \#Tr (\emph{PathSet}/\emph{SegmSet}) & Mean & Std & Median & $20^{th}$ & $80^{th}$ & Max \\
\hline
DT~\cite{oda21} & SL & 5/0 & 342.4 & 236.6 & 308.0 & 156.0 & 518.0 & 910.0 \\
\hline
Zhang et al.~\cite{zhang18} & RL & 5/0 & 496.9 & 250.6 & 441.8 & 340.3 & 607.5 & 1153.7 \\
\hline
Ours (p) & RL & 5/0 & 1333.0 & 515.6 & 1143.3 & 929.9 & 1873.1 & 2456.7 \\
\hhline{=========}
TSP~\cite{shin22_spie} & Graph & 0/20 & 810.0 & 193.6 & 837.0 & 582.0 & 1060.0 & 1162.0 \\
\hline
Ours (s) & RL & 0/20 & 1267.3 & 447.5 & 1243.3 & 1011.9 & 1604.5 & 2370.1 \\
\hline
Ours (s) & RL & 0/25 & 1393.0 & 478.4 & 1302.9 & 1119.4 & 1925.4 & \bf{2532.8} \\
\hhline{=========}
Ours (p+s) & RL & 5/20 & \bf{1519.0} & 500.0 & \bf{1317.2} & \bf{1168.7} & \bf{2183.6} & 2394.0 \\
\hline
Ours (p+s) w/o wall & RL & 5/20 & 1050.0 & 369.4 & 1007.5 & 668.7 & 1523.9 & 1567.2 \\
\end{tabular}
\end{scriptsize}
\end{table}

Table~\ref{tab:quan_res} shows quantitative results of the proposed method and comparable methods. We first provide the performance of the methods that use \emph{PathSet} for training. Since the GT path is available for the entire small bowel in this set, we used the full supervision to train the network proposed in \cite{oda21}, which was originally trained using sparse annotation. Nevertheless, it presented a difficulty on predicting the precise centerline DT for our dataset, which is the key of their method. Using the reward function based on the Euclidean distance to the GT path, `Zhang et al.~\cite{zhang18}', which was originally proposed for the aorta, had trouble with learning a successful policy for the small bowel as well. Our method using the same training set showed a better performance.

The next group of methods are ones that use \emph{SegmSet}. The `TSP' method of \cite{shin22_spie} entails small bowel segmentation, thus the set can be used to train a segmentation network. The proposed method performed reasonably when trained using only \emph{SegmSet}, implying our formulation is operational even without the GT path. By expanding the set a little, `Ours (s) (0/25)' showed a slightly better performance than `Ours (p)'. Considering the respective annotation costs for \emph{PathSet} and \emph{SegmSet}, which were a full day versus a couple of hours per scan, it could be a possible replacement.

Finally, the full method `Ours (p+s)', which uses the both sets for training, performed the best. We can see that including a small number of scans from \emph{PathSet} for training helps increase the performance more by comparing with `Ours (s) (0/25)' again. The use of the wall-related components, namely, the wall input patch and wall-based penalty, are important in the proposed method. It is highlighted when those were eliminated, `Ours (p+s) w/o wall'.

\subsection{Qualitative Evaluation}

\begin{figure}[t]
	\centering
	\begin{minipage}{1\textwidth}
        \subfloat{\includegraphics[width = 0.2\textwidth]{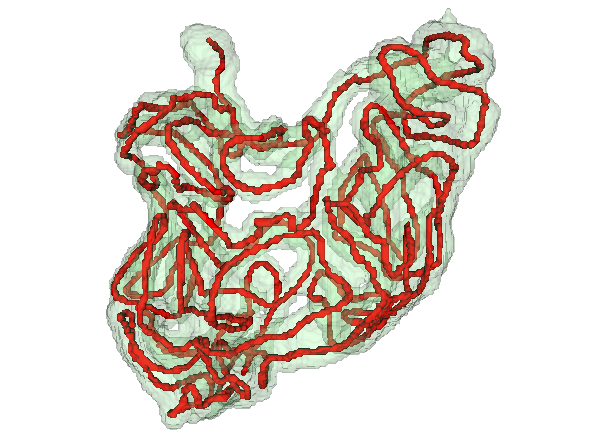}}
	    \subfloat{\includegraphics[width = 0.2\textwidth]{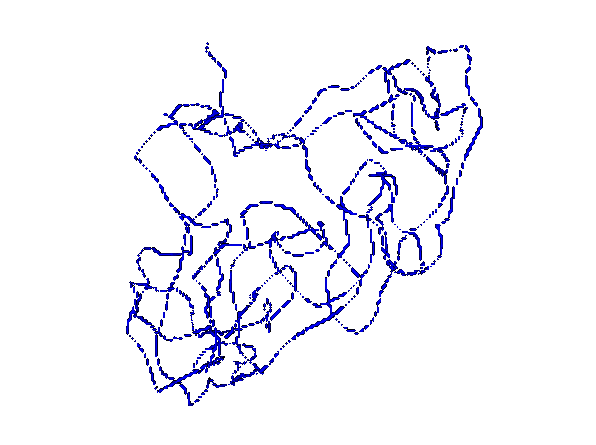}}
	    \subfloat{\includegraphics[width = 0.2\textwidth]{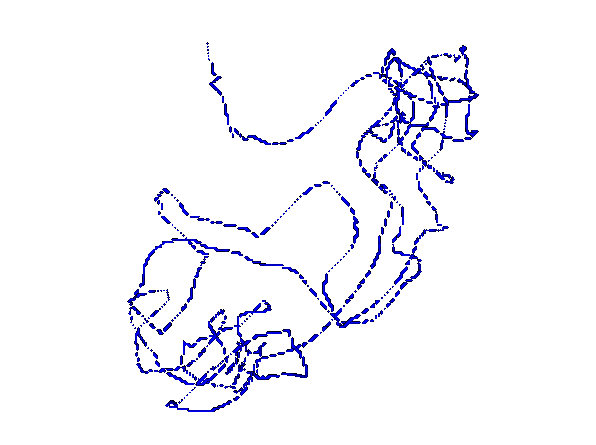}}
	    \subfloat{\includegraphics[width = 0.2\textwidth]{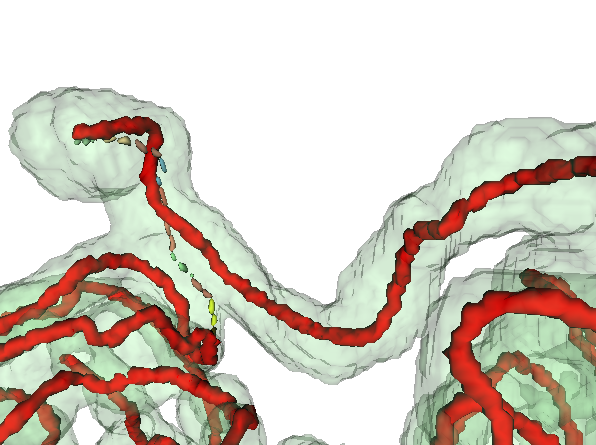}}
	    \subfloat{\includegraphics[width = 0.2\textwidth]{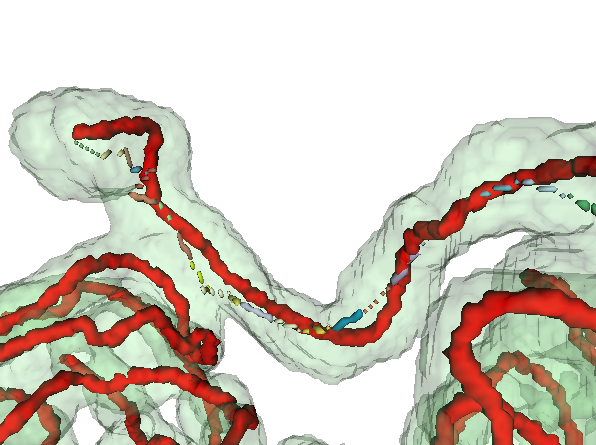}}
    \end{minipage}
    \begin{minipage}{1\textwidth}
        \vspace{-0.4cm}
        \subfloat{\includegraphics[width = 0.2\textwidth]{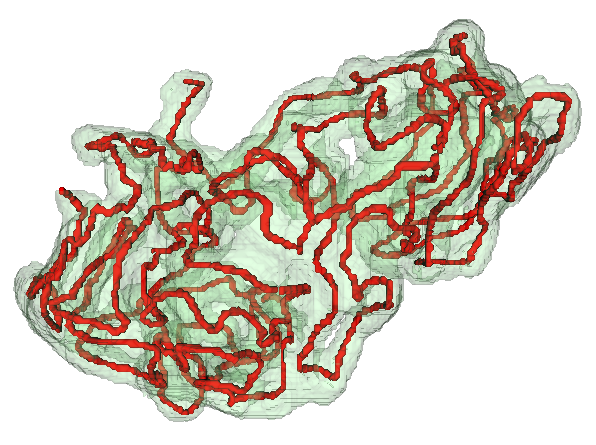}}
	    \subfloat{\includegraphics[width = 0.2\textwidth]{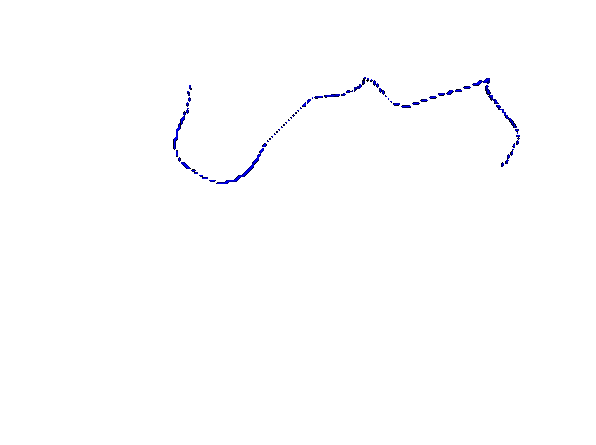}}
	    \subfloat{\includegraphics[width = 0.2\textwidth]{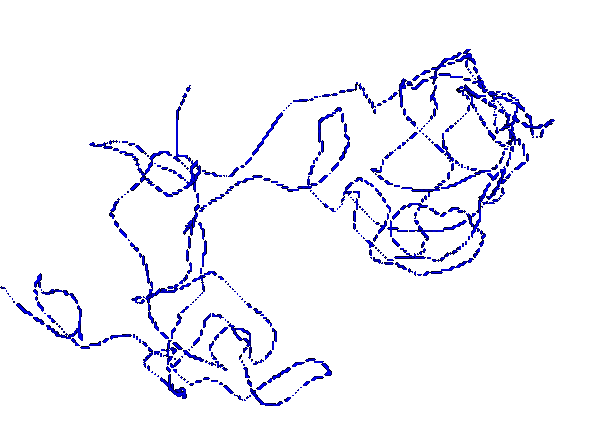}}
	    \subfloat{\includegraphics[width = 0.2\textwidth]{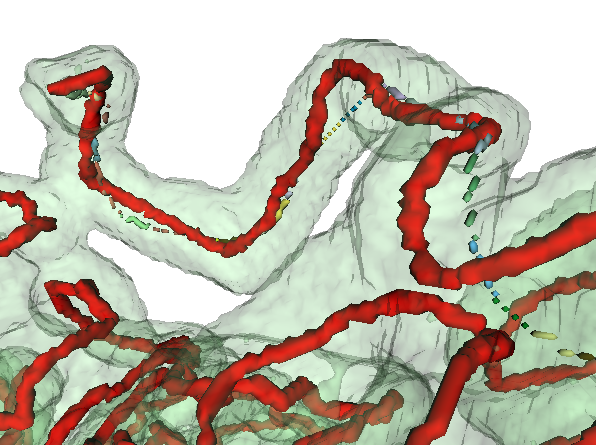}}
	    \subfloat{\includegraphics[width = 0.2\textwidth]{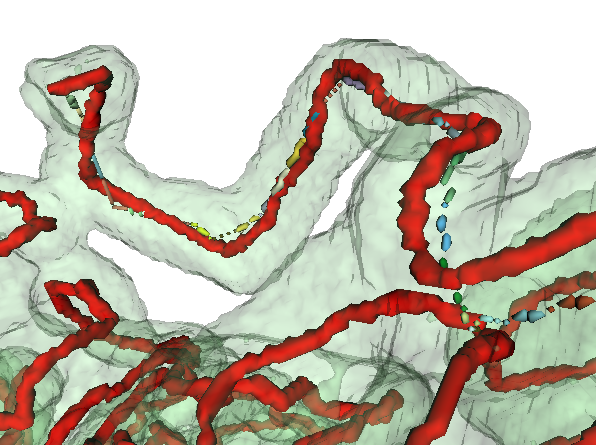}}
    \end{minipage}
    \begin{minipage}{1\textwidth}
        \vspace{-0.4cm}
        \subfloat{\includegraphics[width = 0.2\textwidth]{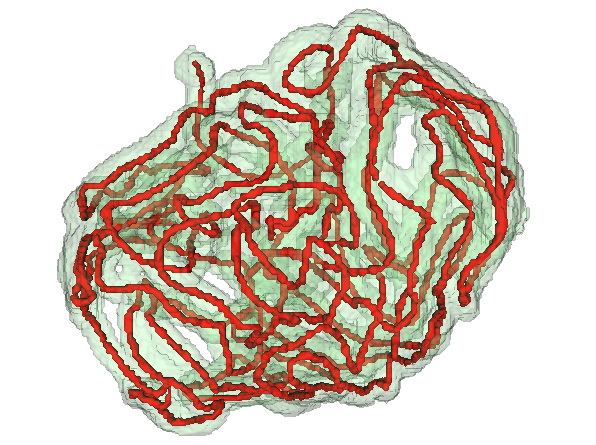}}
	    \subfloat{\includegraphics[width = 0.2\textwidth]{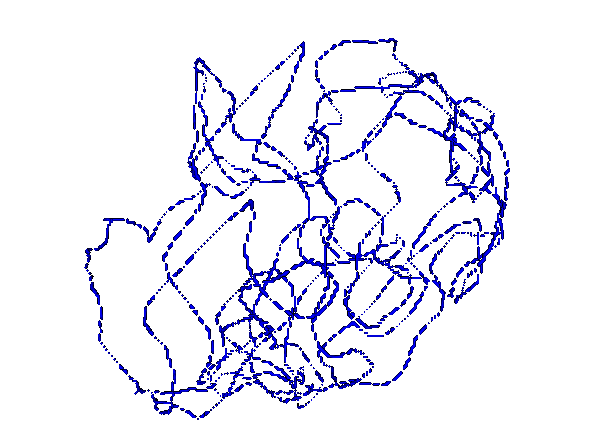}}
	    \subfloat{\includegraphics[width = 0.2\textwidth]{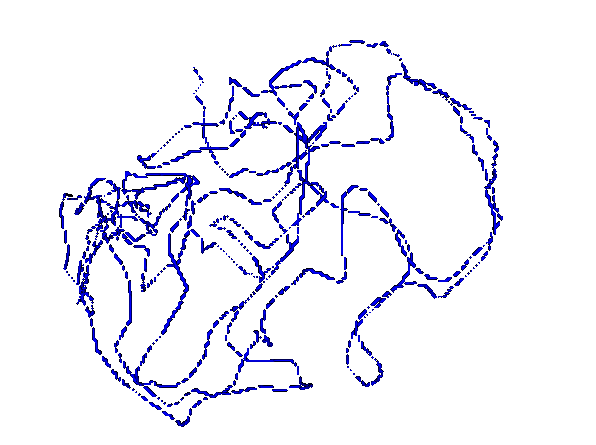}}
	    \subfloat{\includegraphics[width = 0.2\textwidth]{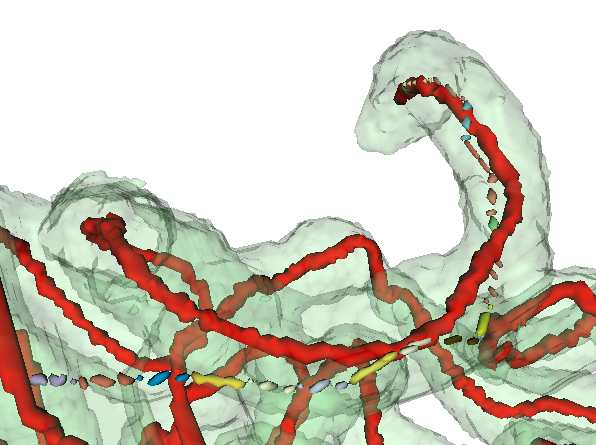}}
	    \subfloat{\includegraphics[width = 0.2\textwidth]{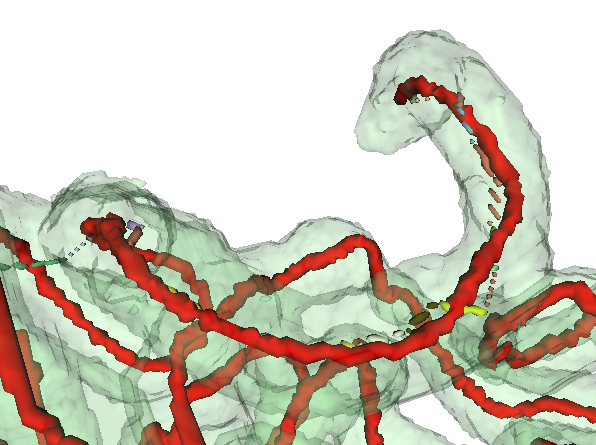}}
    \end{minipage}
	\caption{Example path tracking results. Each row represents different cases. The columns, from left, represent GT path (red) and segmentation (green), result corresponding to `Ours (p)' in Table~\ref{tab:quan_res}, result corresponding to `Ours (p+s)' in Table~\ref{tab:quan_res}, a part of the tracked path corresponding to the second and third columns, respectively. In the last two columns, only a selected local path is shown to highlight the difference, and each step movement is drawn with different colors.}
	\label{fig:qual_res}
\end{figure}

Figure~\ref{fig:qual_res} shows example path tracking results. The result of our full method, which utilizes both of \emph{PathSet} and \emph{SegmSet}, is compared to that of using only \emph{PathSet}. The use of an easier-to-acquire dataset, \emph{SegmSet}, helped achieve a better tracker in a situation where acquiring the GT path in a larger scale is infeasible due to the exceptionally high annotation cost. Note that in the first row, the result of `Ours (p)' has more tracked path overall, but it is achieved by crossing the wall at the first contact area it visits. The full method showed less crossing, resulting in an increase in the maximum length metric.

\section{Conclusion}
We have presented a novel DRL method for small bowel path tracking, which can learn even from scans that are without the GT path. The experimental results showed that it is possible to train a reasonable tracker without using the GT path, and that utilizing those weakly annotated scans together with ones having the GT path can render a better tracker than using either of them alone. Considering the annotation difficulty of the GT path, the proposed method could reduce the annotation cost required for training. The tracked path can provide better information on the small bowel structure than segmentation. Especially, it could be useful for image-guided intervention where a device approaches according to the identified structure.

\subsubsection*{Acknowledgments.} We thank Dr. James Gulley for patient referral and for providing access to CT scans. This research was supported by the Intramural Research Program of the National Institutes of Health, Clinical Center. The research used the high performance computing facilities of the NIH Biowulf cluster.

%
%

%
%
%
%


\title{Deep Reinforcement Learning for\\Small Bowel Path Tracking using\\Different Types of Annotations:\\Supplementary Material}
\titlerunning{Deep Reinforcement Learning for Small Bowel Path Tracking}
%
\author{Seung Yeon Shin \and
	Ronald M. Summers}
\authorrunning{S.Y. Shin et al.}
%
\institute{Imaging Biomarkers and Computer-Aided Diagnosis Laboratory, Radiology and Imaging Sciences, Clinical Center, National Institutes of Health, Bethesda, MD, USA}
\maketitle              
\begin{algorithm}[h]
	\caption{Training for small bowel path tracking}
	\label{alg:train_alg}
	\begin{algorithmic}[1]
		\Require{The number of episodes from \emph{PathSet} $N_p$, the number of episodes from \emph{SegmSet} $N_s$, the maximum time step per episode T}
		\Ensure{Actor network $A(w^{a})$, critic network $C(w^{c})$}
		\State Initialize the actor/critic networks $A(w^{a})$, $C(w^{c})$
		\Repeat
		\For{i=1 to $N_p+N_s$}
		\If{$i<=N_p$}
		\State Sample an image from \emph{PathSet}
		\Else
		\State Sample an image from \emph{SegmSet}
		\EndIf
		\State Initialize the state for the actor network $s_a$
		\State Initialize the state for the critic network $s_c$
		\For{t=1 to T}
		\State Sample an action $a_t$ from predicted beta distributions using $A(s_a|w^{a})$
		\State Keep $a_t$'s probability of being sampled $\pi_{w^{a}}(a_t|s_a)$
		\State Execute $a_t$ and get new states $s_a'$, $s_c'$, reward $r_t$, and termination
		\State Predict the value $v_t=C(s_c|w^{c})$
		\State Store ($s_a$, $s_c$, $a_t$, $r_t$, $v_t$, $\pi_{w^{a}}(a_t|s_a)$, termination)
		\If{termination}
		\State Break
		\EndIf
		\State $s_a \leftarrow s_a'$, $s_c \leftarrow s_c'$
		\EndFor
		\EndFor
		\State Update the actor/critic networks $A(w^{a})$, $C(w^{c})$ using PPO
		\Until{$A(w^{a})$ and $C(w^{c})$ converge}
	\end{algorithmic}
\end{algorithm}

\end{document}